\documentclass{nature}
\usepackage{graphicx,xcolor,ulem}
\makeatletter
\let\saved@includegraphics\includegraphics
\AtBeginDocument{\let\includegraphics\saved@includegraphics}
\renewenvironment*{figure}{\@float{figure}}{\end@float}
\makeatother
\title{A magnetically-driven equatorial jet in Europa's ocean}

\author{Christophe Gissinger$^{1}$, Ludovic Petitdemange$^2$ }

\newcommand{\be}{\begin{equation}}
\newcommand{\ee}{\end{equation}}

\renewcommand{\vec}[1]{{\bf #1}}

\renewcommand{\Re}{{\rm {I\hskip -0.55mm Re}}}

\begin{document}

\maketitle

\begin{affiliations}
 \item Laboratoire de Physique de l'Ecole Normale Superieure, ENS, Universit\'e PSL, CNRS, 24 rue Lhomond, 75005 Paris, France (corresponding author)
 \item LERMA, CNRS, Paris, France
\end{affiliations}

\begin{abstract}
 During recent decades, data from space missions have provided strong evidence of deep liquid oceans underneath a thin outer icy crust on several moons of Jupiter~\cite{Neubauer1998a, Khurana1998}, particularly Europa~\cite{Roth2014, Sparks2017}. 
  But these observations have also raised many unanswered questions regarding the oceanic motions generated under the ice, or the mechanisms leading to the geological features observed on Europa~\cite{Pappalardo1996,Pappalardo1999}. By means of direct numerical simulations of Europa's interior, we show here that Jupiter's magnetic field generates a retrograde oceanic jet at the equator, which may influence the global dynamics of Europa's ocean and contribute to the formation of some of its surface features by applying a unidirectional torque on Europa's ice shell.
\end{abstract}

Whereas both radiogenic and  tidal heating~\cite{Ross1986, Spohn2003} produce the energy dissipation necessary to the melting of the ice~\cite{Bills2005,Tyler2008}, motions in the ocean underneath the Jovian moons are believed to be generated through vigorous thermal convection~\cite{Soderlund2013}, hydrothermal plumes~\cite{Goodman2004,Goodman2012} or double-diffusion convection~\cite{Vance2005}. Such flows certainly play a dominant role, but may fail at explaining some of the observations if considered alone~\cite{Spohn2003}, strongly suggesting the presence of an additional physical mechanism in these oceans.
Because the magnetic dipole axis is tilted by about $10^{o}$  with the rotation axis of the gaseous giant, Jupiter's moons also experience a time-varying magnetic field  with a rotation rate $\omega$, inducing electrical currents in the oceanic salty water~\cite{Colburn1985}.


Here, we argue that as long as the phase lag between the induced field and the Jovian one is non-zero, these induced currents naturally combine with the magnetic field to generate a Lorentz force, leading to a weak magnetohydrodynamic (MHD) process that might play a significant role on the global dynamics of the ocean. We therefore model Europa's interior as a spherical shell (mean radius $R=(R_i+R_E)/2$, thickness $h=R_E-R_i$) of salty water (electrical conductivity $\sigma$ and kinematic viscosity $\nu$) confined between an inner mantle of silicate rocks (radius $R_i$) and an outer layer (radius $R_E$) of ice crust (see our Method section for a definition of the control parameters). 
 We specifically model Europa here, but our results should apply equally to subsurface oceans found in other Jovian moons.  In order to focus on the MHD process, we first present simulations in which thermal buoyancy is neglected. In the last part of this paper and in Supplementary Material, we show that buoyancy only weakly modifies the magnetically-driven jet, but remains crucial to get a full picture of the ocean dynamics.  
 
\begin{figure}[htb!]
\begin{center}
 \includegraphics[width=0.95\textwidth]{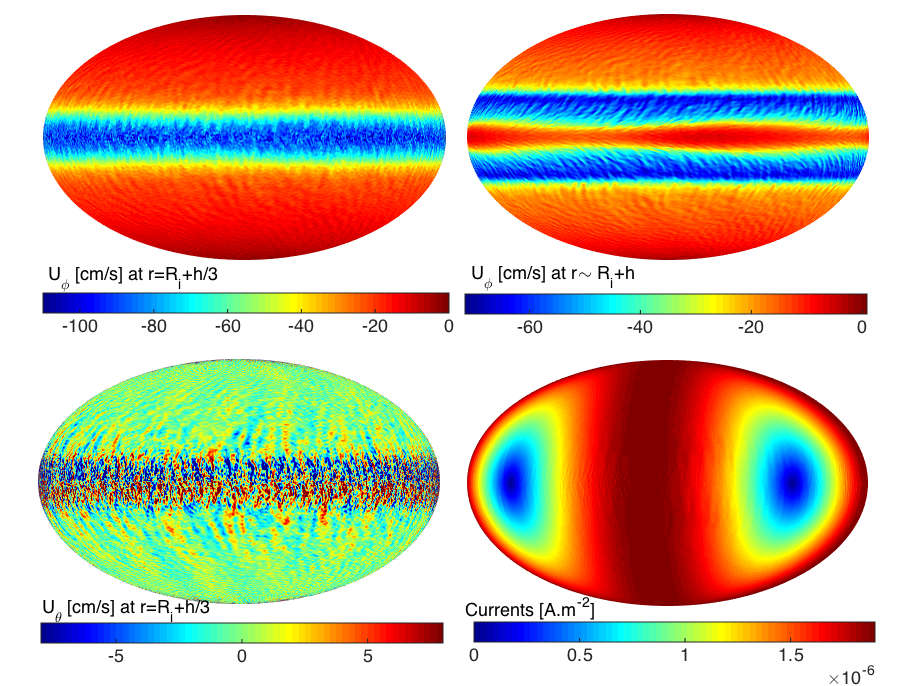}
\caption{{\bf Velocity field and ohmic currents.} Snapshot of the azimuthal $U_\phi$ and polar $U_\theta$ components of the velocity field (a-c) and of the ohmic currents $\sqrt{|J|^2}$ (d). The fields are shown in the $\phi,\theta$ plane (elliptical projection) for one of our global models of Europa, $Ek=10^{-7}$, $\Lambda=10^{-1}$, $Pm=10^{-4}$, $h=100$km, corresponding to the highest possible value for $N$ (see text). $R_i$ is the radius of the inner sphere, and $h$ is the ocean's thickness. Panel (b) shows that due to geostrophic constrain, the flow exhibits a more complex structure close to the ice shell.}
  \label{4spheres}
 \end{center}
\end{figure}

Fig.1 shows that the rotation of Jupiter's magnetic field induces a planetary scale recirculation, and generates strong upward and downward turbulent motions at the equator (Fig. 1c). But the most striking feature is the generation  of a powerful oceanic jet propagating westward (Fig. 1a and 1b), and localized in the moon's equatorial region. The Jupiter-Europa system can therefore be regarded as a gigantic induction electromagnetic pump, in which the salty water of the subsurface ocean is electromagnetically pumped at a mean velocity ${\bf \overline{U}}$ by the variations of the Jovian magnetic field travelling in the horizontal plane at speed $c=\omega R\sim 230$m.s$^{-1}$. This can be easily understood from the induction equation:

\begin{equation}
\frac{\partial {\bf B}}{\partial t}={\bf \nabla\times}\left({\bf U\times B}\right)+\frac{1}{\mu_0\sigma}{\bf \nabla^2 B}
\label{ind}
\end{equation}

which governs the evolution of the magnetic field inside Europa's ocean. If one assumes that the field can be written $(B^J_r{\bf e_r}+B^J_\phi{\bf e_\phi}) e^{i(\phi-\omega t)}$ and that the induced currents  ${\bf J}=\sigma(U(r)-c)B_r{\bf e_\theta}$ are also travelling waves, it follows that the mean Lorentz force acting on the ocean can be written (see Methods):

 \begin{equation}
{\overline {\bf F}}= \frac{\sigma B_0^2R^2(c-\overline{U})}{h^2\left(8+2\mu_0^2\sigma^2R^2(c-\overline{U})^2\right)}{\bf e_\varphi}
\label{F_L}
\end{equation}

This expression of the Lorentz force is well known in the context of electromagnetic pumps~\cite{Gailitis76, Reddy2018}, and describes how a small driving of the ocean is produced as long as $\sigma$ is finite. An important feature is the phase lag $\phi_l$ between the Jovian field $B_0$ and the induced one, which controls the torque applied on the ocean. Our simulations span a large range of values of $\phi_l$, but cases corresponding to Jovian moons  exhibit phase lags between $1$ and $3$ degrees (see supplementary Fig. 3). Accordingly, the Lorentz force is very small, and these magnetically-driven jets are expected to be very weak compared to the velocity of the Jovian field. In other words, Jovian moons are {\it inefficient} induction pumps producing induced fields almost identical to the ones predicted by non-MHD studies.

Note that the dimensioned values indicated in Fig.1 are specific to the set of parameters used here, still very far from the one relevant to real moons. Thus, in order to estimate the value of the Lorentz force $F$ acting on Europa's ocean, we next compare our DNS to observations. Following previous studies, we  use Galileo measurements of Europa's induced magnetic field to constrain the value of the ocean's electrical conductivity. To wit, we extensively explored our parameter space and compared our results to the spacecraft mission's flyby $E14$. As shown in Fig.2, we found that the best fits are obtained for $\sigma$ in the range $0.3-3$ S.m$^{-1}$ if the ocean depth is $h=147$km. This value, corresponding to $Rm=\mu_0\sigma ch$ comprised between $8$ and $80$, is in good agreement with previous predictions~\cite{Schilling2007} and implies a salinity comparable to terrestrial oceans~\cite{Hand2007}. 

\begin{figure}[htb!]
\begin{center}
\includegraphics[width=0.9\textwidth]{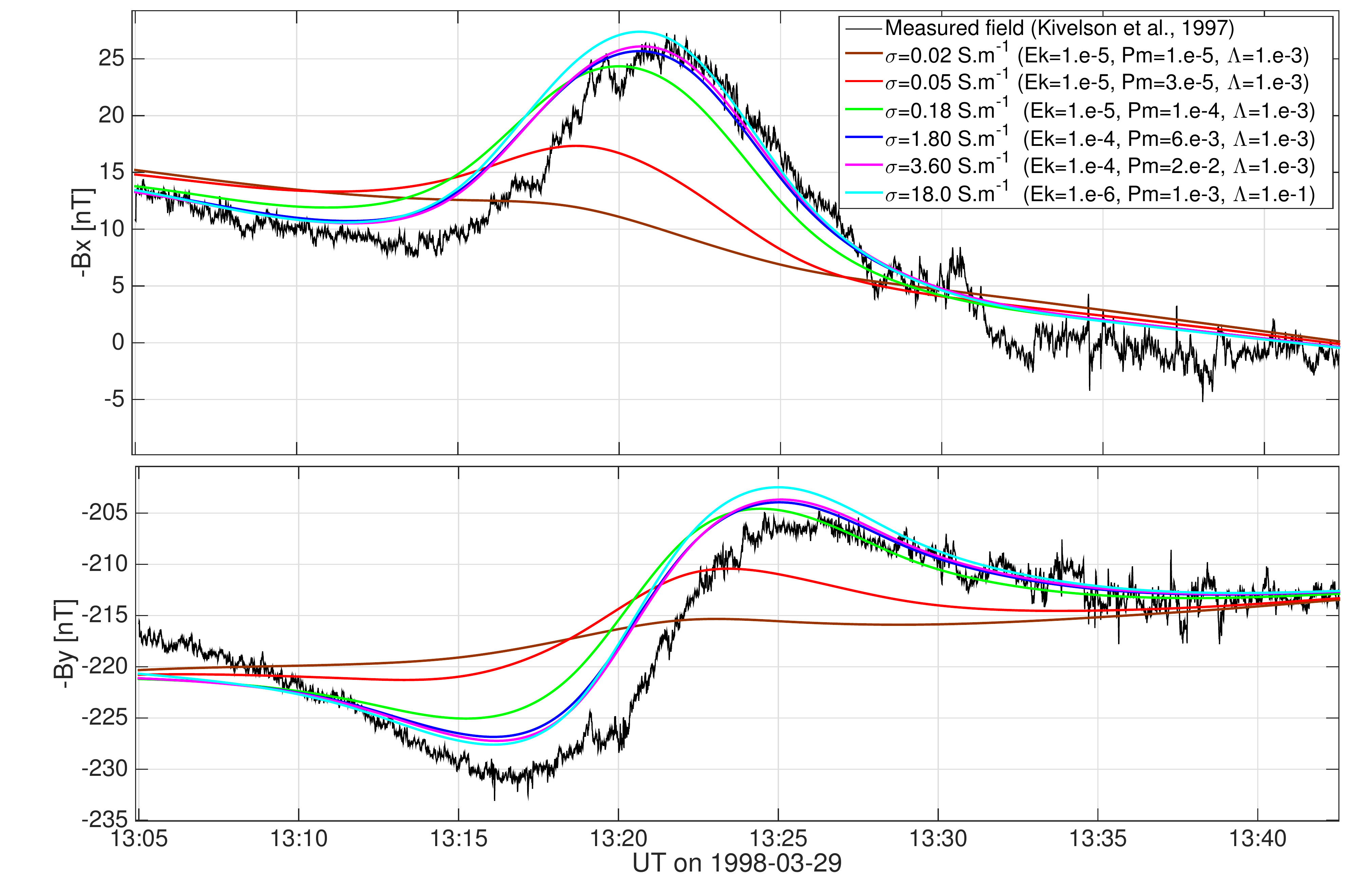}
\caption{{\bf Galileo magnetometer data in the EPhiO coordinates (Europa-centered with $x$ along the direction of corotation, $y$ radially inward toward Jupiter, and $z$ parallel to Jupiter's rotation axis), compared to results from DNS using $h=147$km}. The black curves show time evolution of magnetic field components $B_x$ (top) and $B_y$ (bottom) recorded by Galileo during the trajectory {\it E14}, while other solid lines correspond to our DNS with various values of the electrical  conductivities $\sigma=2\times 10^{-2}; 5\times 10^{-2}; 1,8\times 10^{-1}; 1.8 ; 3.6 $ and $18$ S.m$^{-1}$. Best fits are obtained for electrical conductivities between $0.3$ and $3$ S.m$^{-1}$.
}
\label{fig:galileo}
\end{center}
\end{figure}

Quantitative predictions clearly require to identify which term balances the time-averaged Lorentz force $\overline{F}$ at large scale. Upon azimuthal averaging, thermal buoyancy does not contribute much, and the central question is to know which of the viscous force or the nonlinear term balances $\overline{F}$. As shown in Supplementary material II, a purely viscous balance (ignoring non-linear terms and global rotation) leads to velocity of the oceanic jet such that:

\begin{equation}
\overline{U}\propto\frac{B_0^2}{\mu_0^2\rho\nu\sigma \omega R_E},
\label{theory}
\end{equation}

meaning that $Q=\overline{U}/c$ depends on one dimensionless number only, $N=B_0^2/(\mu_0^2\rho\nu\sigma c^2)$. This formula predicts a very strong equatorial jet of a few $km/h$ when applied to Europa's parameters. On the other hand, one may rather expect a fully inertial regime (ignoring the viscous term), in which a significant part of the injected power per area $P_{ohm}\sim2B_0^2/\mu_0^3\sigma^2ch^2$ is evacuated through turbulent dissipation. In this case, boundary-layer theory predicts $\gamma P_{ohm}\sim\rho C_DU^3$, leading to a different scaling for the jet velocity $\overline{U}\sim (2\gamma B_0^2/\rho C_D\mu_0^3\sigma^2ch^2)^{1/3}$, where $C_D$ is the drag coefficient and $\gamma$ is the ratio between viscous and ohmic dissipation. This prediction rather leads to jet velocities of a few mm.s$^{-1}$. 
\begin{figure}[htb!]
\begin{center}
\includegraphics[width=0.95\textwidth]{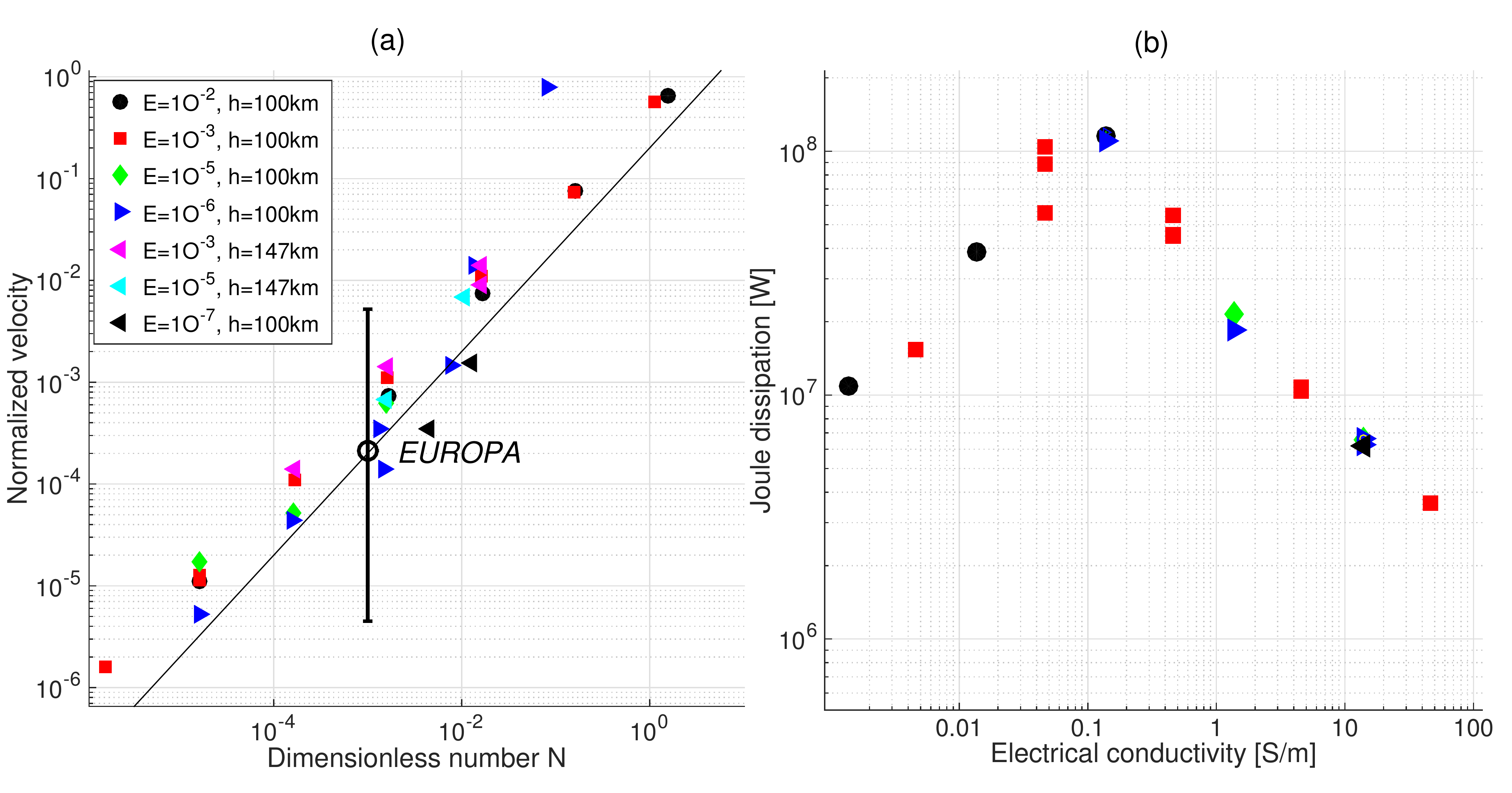}
\caption{{\bf Magnitude of the oceanic jet.} {\bf a}.Predictive scaling law for the magnitude of the oceanic jet. DNS computed for various values of the Ekman number $Ek$ and the ocean thickness $h$ collapse on the law $\overline{U}/c\propto N$, with $N=B_0^2/(\mu_0^2\rho\nu\sigma c^2)$ (solid black line). Estimate of $N$ leads to an oceanic jet of a few $cm/s$ is for Europa (black empty circle), error bars indicating maximal and minimal bounds on the value of eddy viscosity. {\bf b}, Corresponding Joule dissipation versus the electrical conductivity of the ocean.}
\label{fig:predictions}
\end{center}
\end{figure}

Finally, an intermediate approach is to assume an eddy viscosity $\nu_t=\nu+C_D\overline{U}h$ in the laminar formula (\ref{theory}). Fig.3a shows a fairly good rescaling of all simulations when $Q$ is plotted as a function of $N$ using such an eddy viscosity. This suggests that molecular viscosity probably provides unrealistically high velocities. On the other hand, because the global rotation produces a weakening of the turbulence intensity due to the two-dimensionalization of the large-scale flow~\cite{Campagne2016}, an eddy viscosity much smaller than the one observed in non-rotating turbulence is expected. In consequence, the viscous and turbulent scalings discussed above might be regarded as maximal and minimal bounds on the jet's magnitude.  Under the assumption that Jovian moons lie between these two regimes, our simulations therefore predict that Europa have the most powerful jet, with a time-averaged azimuthal flow possibly reaching a few cm.s$^{-1}$, while Ganymede should exhibit a few mm$.s^{-1}$ and Callisto a nearly negligible jet ($\overline{U}< 1$ mm$.s^{-1}$) 

Because tidal and radiogenic heating generate hydrothermal plumes with velocities estimated at several cm.s$^{-1}$, the electromagnetically-driven jet described here could very well be negligible compared to those thermally-driven flow~\cite{Goodman2004,Goodman2012}, especially because simple estimates of the Lorentz force gives very small values ($F_B\sim 10^{-13}$ N.m$^{-3}$). To address this question, we now report in Fig.4 a simulation including thermal buoyancy, in which both $N$ and the convective Rossby number are such that the magnitudes of the magnetically-driven jet and the thermally-driven flows are similar to what presumably occurs in Europa's ocean.
As expected~\cite{Goodman2012}, geostrophic thermal plumes strongly dominate at small scale, with velocities around $10$ cm.s$^{-1}$ and typical diameters of $30$ km. The convective Rossby number $Ro\sim3\times10^{-2}$ being too small to generate a significant zonal wind, the same simulation with no magnetic forcing (not shown here) displays no large scale component of the azimuthal velocity field at the equator. Europa's zonal flow shown in Fig.4 is therefore entirely due to Jupiter's field. Because this magnetically-driven jet is the main contribution to the axisymmetric time-averaged azimuthal velocity field ($\overline{U_\phi}\sim 2 $cm.s$^{-1}$), it is not suppressed by the vigorous buoyancy force, even with such a small Lorentz force. Note that a more complex situation arises if the convective Rossby number is of order $1$, due to the generation of a strong thermal wind~\cite{Soderlund2013} (see our discussion in Methods and Sup. Fig.1).

 We now discuss some consequences of the existence of a magnetically-driven jet on Europa. First, even at 1mm.s$^{-1}$, such a zonal flow might play a central role in the potential development of life, by providing a time-independent westward transport of radiolytically-produced oxidants and other biologically useful substances \cite{Greenberg2010}. Second, the inefficiency of the magnetic pumping ($U\ll c$) yields modest ohmic dissipation, expected to be around $10^7$-$10^8$ W on Europa ( for $Rm\sim 8-80$, see Fig.3b). Radiogenic ($10^{11}$ W) or solid body tidal heating ($10^{12}$ W) therefore represent a much larger contribution. Note however that skin effect concentrates ohmic heating inside a very thin layer close to the ice in the polar region (see Fig.4, right), such that the ohmic dissipation can locally reach between $10^{-2}$ and $10^{-1}$ mW.m$^{-2}$. This represents a non-negligible fraction of the obliquity tidal heating $3\times 10^{-1}$ mW.m$^{-2}$ estimated in the ocean~\cite{Tyler2008}, which also peaks near Europa's poles  where water plumes are preferentially generated~\cite{Sparks2016}.   

More important, contrary to geostrophic buoyancy plumes or time-periodic tidally-driven flows, the magnetically-driven jet  provides a constant unidirectional torque on Europa's ice shell. This new azimuthal force may have a direct influence on the reorientation of Europa's ice shell in the long term. It has been proposed that reorientation could take the form of a non-synchronous rotation produced by tidal torques~\cite{Greenberg1984}, or true polar wander in the case of an asymmetric tidal heating in the ice shell~\cite{Ojakangas1989}. Non-synchronous rotation is also regularly invoked to explain some of the geological features observed at the surface of Europa \cite{Helfenstein1985}, and some evidences of reorientation of the ice shell were reported~\cite{Geissler1998, Schenk2008}. The question of whether or not Europa's ice shell has reoriented in the past involves many complex phenomena (mainly based on tidal forces), but our results suggest that the magnetic field of Jupiter, by generating a net westward motion in the ocean, can influence any possible non-synchronous rotation of Europa. Taking into account this new effect may therefore shed a new light on the formation of Europa's global system of lineaments.
  
\begin{figure}[htb!]
\begin{center}
\includegraphics[width=1.\textwidth]{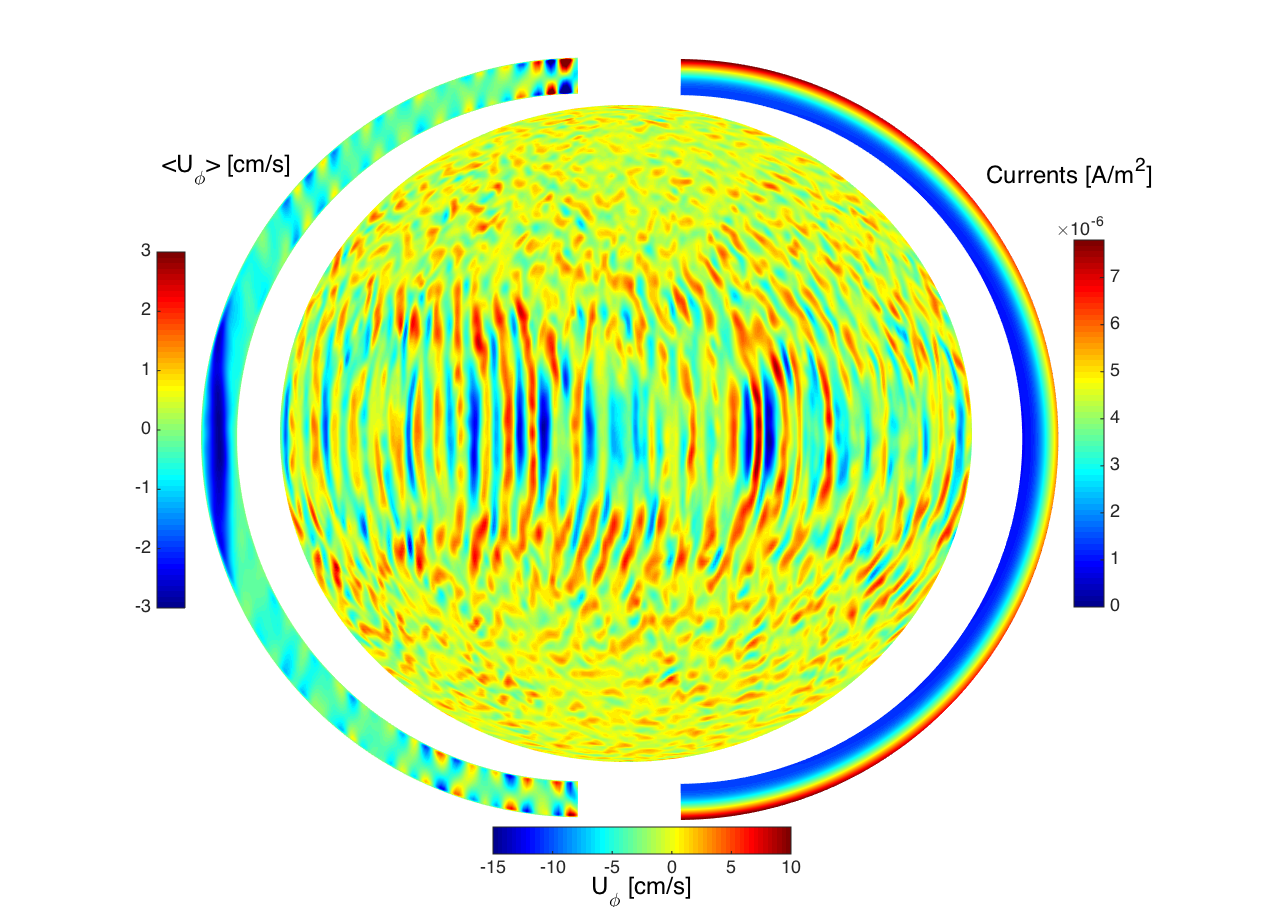}
\caption{{\bf Snapshot of the instantaneous azimuthal velocity field at $r=R_i+h/2$ from the seafloor, for Ekman number $E=10^{-5}$, $Pr=12$, $Pm=10^{-3}$, $\Lambda=10^{-3}$ and $Ra=10^9$}. This set of parameters corresponds to $N=2\times 10^{-4}$, a value expected for Europa in presence of eddy viscosity. Left and right panels show respectively $\phi$-averaged velocity field $U_\phi$ and electrical currents $\sqrt{|J|^2}$. Both the magnetically-driven zonal flow and thermal plumes are geostrophic.  }
\label{fig:rtheta}
\end{center}
\end{figure}

Note that a more complete modeling of subsurface oceans would certainly require to describe tidal effects or topological features of the ice, and to reach more realistic Ekman numbers. In this perspective, the upcoming space missions {\it JUICE} (JUpiter ICy moons Explorer)~\cite{Juice} and {\it Europa Clipper}~\cite{Clipper} should help constraining the future models, for instance by providing more precise estimates on the phase lag $\phi_l$ of the induced field. On the other hand, global DNS of Europa's interior, by clarifying the dynamics generated below the ice, may already be helpful in the design of these future spacecraft missions.


\bibliographystyle{naturemag}
\bibliography{sample}

\begin{addendum}
\item[Acknowledgment] The GO-J-MAG-3-RDR-HIGHRES-V1.02 data set was obtained from the Planetary Data System (PDS). This work was granted access to the HPC resources of MesoPSL financed by the Region Ile de France and the project Equip@Meso (reference ANR-10-EQPX-29-01) of the programme Investissements d'Avenir supervised by the Agence Nationale pour la Recherche.
\item[Author contributions] C.G. conceived the presented idea and developed the theory, L.P. performed the numerical simulations. C.G. and L.P. performed the analysis of the results and contributed to the final manuscript. The authors declare that they have no
competing financial interests.
\item[Correspondence] Correspondence and requests for materials
should be addressed to C. Gissinger~(email: christophe.gissinger@lps.ens.fr).
  \end{addendum}

\newpage


\begin{methods}

{\bf  1. Numerical modeling}\\

We simulate Europa's interior as  an electrically  conducting fluid confined in a rotating spherical shell subject to an oscillatory background field $\vec{B_0}$ generated by the host planet Jupiter. In space, such a field must satisfy $\vec{\nabla}\times\vec{B}_0=0$, and mainly consist of a dipole field with a stationary axial dipolar component and an oscillatory equatorial component. We use a spherical coordinate system, with $r$ the radial distance from Europa's center, $\theta$ the polar angle measured from the North pole and $\phi$ the azimuth angle counted positive along the eastward prograde direction. In this system, the dimensionless non-stationary magnetic field of Jupiter $\tilde{B_0}$ writes:

$$
 \tilde{B_r^0}=  \frac{2\sin\theta\cos(\omega \tilde{t} -\phi)}{\tilde{r_j}^3} - \frac{3\sin\theta\cos\phi\cos\omega \tilde{t}}{\tilde{r_j}^5} 
+\frac{3\gamma \tilde{r}}{\tilde{r_j}^5}\left[  \sin\omega \tilde{t} -\sin^2\theta\sin\phi\cos(\omega \tilde{t} -\phi)  \right] \\
$$
$$
\tilde{B_\theta^0}= \frac{2\cos\theta\cos(\omega \tilde{t} -\phi)}{\tilde{r_j}^3} - \frac{3\cos\theta\cos\phi\cos\omega \tilde{t}}{\tilde{r_j}^5} -\frac{3\gamma \tilde{r}}{\tilde{r_j}^5}\left[  \sin\theta\cos\theta\sin\phi\cos(\omega \tilde{t} -\phi) +\gamma \tilde{r}\cos\theta\cos(\omega \tilde{t}-\phi) \right] \\
$$
$$
\tilde{B_\phi^0}=\frac{2\sin(\omega \tilde{t} -\phi)}{\tilde{r_j}^3} + \frac{3\sin\phi\cos\omega \tilde{t}}{\tilde{r_j}^5} 
+\frac{3\gamma \tilde{r}}{\tilde{r_j}^5}\left[  \sin\theta\cos\omega \tilde{t} -\sin\theta\sin\phi\sin(\omega \tilde{t} -\phi) -\gamma \tilde{r} \sin(\omega \tilde{t} -\phi)  \right] 
$$
where $\omega$ is the dimensionless rotation rate of the field in the framework of Europa, $\tilde{r}$ is the radius normalized by $R_E$, and $\tilde{r_j}$ is given by $\tilde{r_j}=\sqrt{1+\gamma^2 \tilde{r}^2+2\gamma \tilde{r}\sin\theta\sin\phi}$.  The parameter $\gamma=R_E/R=2\times10^{-3}$ is the ratio between Europa's radius and the distance Jupiter-Europa. Such a small value for $\gamma$ corresponds to an almost constant magnetic field rotating in the equatorial plane, although our simulations take into account the small variation of the field due to the fact that $\gamma$ is not exactly zero. In a few simulations (not reported here), we observed that the retrograde jet is slightly smaller when the constant vertical component of the magnetic field is removed, an effect that will be described elsewhere.

In Europa's conducting ocean, the magnetic field can be written as $\vec{B}=\vec{B}^0+\vec{b}$, where $b$ is the induced magnetic field. 
The oceanic saltwater  is an electrically conducting fluid (conductivity $\sigma$) satisfying Maxwell equations:

\begin{equation}
\frac{\partial {\bf B}}{\partial t}=-{\bf \nabla\times} {\bf E} \hspace{2cm} \mu_0{\bf J}={\bf \nabla\times B}
\label{maxwell}
\end{equation}

By combining these equations with the modified Ohm's law ${\bf J}=\sigma({\bf E}+ {\bf u\times B})$, one obtains the induction equation which gives the evolution of the magnetic field and induced currents. When this equation is coupled to the Navier-Stokes equation and the problem is made dimensionless, one finally obtains magnetohydrodynamic (MHD) equations for a rotating conducting Boussinesq fluid:

\begin{equation}
\frac{1}{E}\left(\frac{\partial \bf{\bf u}}{\partial t}+({\bf u}\cdot{\bf \nabla}){\bf u}+2{\bf e_z}\times {\bf u} +{\bf \nabla}\pi\right)=  \Delta {\bf u} \\ \nonumber
    +\frac{\Lambda }{Pm}({\bf \nabla}\times{\bf B})\times{\bf B} \, ,
\label{NSnumeric}
\end{equation}

\begin{equation}
\label{INDnumeric}
\frac{\partial {\bf b}}{\partial t}=\frac{E}{Pm} \Delta {\bf b}+{\bf \nabla}\times\left({\bf u}\times({\bf B_0}+{\bf b})\right)-\frac{\partial {\bf B}_0}{\partial t} \, ,
\end{equation}

\begin{equation}
{\bf \nabla}\cdot{\bf b}= 0 \, , \qquad {\bf \nabla}\cdot{\bf u}= 0 \, ,
\end{equation}
where we introduced the Ekman number (ratio of viscous to Coriolis forces), the Elsasser number (ratio of Lorentz to Coriolis forces) and the magnetic Prandtl number (ratio between the viscosity and the magnetic diffusivity), defined as
\begin{equation}
E=\nu/(\Omega_0 R_E^2)\; , \qquad \Lambda=B_0^2/(\mu_0\rho_0\eta \Omega_0) \; , \qquad Pm=\nu/\eta \; .
\end{equation}

As usual, $\nu$, $\eta=\mu_0\sigma$, $\rho_0$, $\mu_0$, $\sigma$ denote the kinematic viscosity,  the magnetic resistivity,  the density, the magnetic permeability and the electrical conductivity, all assumed to be constant.  $\Omega_0$ denotes the angular rotation of Europa, which serves as reference frame. In the above, we used  the radius of the spherical domain of Europa $R_E$ as unit of length, and $1/\Omega_0$ as unit of time. The ocean's thickness is varied from $100$km to $147$km, which  are the typical values generally discussed in the literature~\cite{Hand2007}. Both the ocean-mantle and ocean-ice boundaries are assumed to be electrically insulating and isothermal. No-slip boundary conditions for the velocity field are used, and internal heat sources that may be generated by tides or libration are neglected.

Alternatively, one may also define the Hartman number $Ha=B_0h\sqrt{\sigma/\rho\nu}$, which compares the Lorentz force to the viscous force, and the magnetic Reynolds number $Rm=\omega R_Eh\mu_0\sigma$, which measures the ratio between induction and diffusion. Supplementary table 1 shows that our typical Ekman numbers are many order of magnitudes larger than the value expected in Jovian moons, but Europa's Hartmann and  magnetic Reynolds numbers are correctly reproduced by our simulations. This means that our DNS correctly describe the electromagnetic processes occurring in the ocean, but probably underestimate the effect of global rotation on the flow structure. This underestimate of the global rotation also forces us to use larger values of the Elsasser number. Similarly, the weak magnetic Prandtl number used in the simulations implies that the small scale turbulence of the ocean is not fully resolved, although large values of $Pm$ may also reflects effects due to turbulent diffusivities. Finally, our Rayleigh numbers are chosen to select a convective Rossby number $Ro=\sqrt{RaE^2/Pr}$ in the range estimated for Europa\cite{Soderlund2013}. This ensures that the size and velocities of thermal plumes are correctly reproduced, and that Europa's ratio between the Lorentz and Buoyancy forces is correctly reproduced. Interestingly, both the magnetic Ekman number $E_m=E/Pm$ which appears in the induction equation (\ref{INDnumeric}) and the modified Elsasser parameter $\Lambda_m=\Lambda/Pm$ in equation (\ref{NSnumeric}) are correctly reproduced in the numerical model. This similarity in the force balance may explain why our model successfully reproduce several aspects of Europa despite strong differences in $E$, $Pm$ and $\Lambda$. In terms of dimensional units, apart from viscous and thermal diffusivities which are orders of magnitude too large compared to Europa's ocean, all other parameters are correctly reproduced by the simulations: ocean's thickness $h$ is either $100$ or $147$ km, gravitational acceleration is $g=1,3$m.s$^{-2}$, thermal expansion coefficient $\alpha=3\times10^{-4}$K$^{-1}$, the global rotation rate is $\Omega = 2,1\times10^{-5}$ s$^{-1}$, the magnetic field rotation rate is $\omega=1,6\times 10^{-4}$s$^{-1}$ and the electrical conductivity is $\sigma\sim 1$S.m$^{-1}$.

The numerical benchmarked$^{31}$ solver used to compute MHD equations is the semi-spectral code PaRoDy$^{32}$. The code uses a poloidal/toroidal expansion and a pseudo-spectral spherical harmonic expansion, while the radial discretization is based on finite differences on a stretched grid (allowing for a parallelization by radial domain decomposition). The time integration is performed using a Crank-Nicholson scheme for diffusion terms and an Adams-Bashforth scheme for other terms. For the lowest Ekman numbers reported in this paper, we use 288 points in the radial direction and 305 spherical harmonic modes, corresponding to 460 points in the $\theta$ direction and 1024 points in the azimuthal direction $\phi$.

\noindent{\bf  2. Galileo measurements}

If we assume that the ice crust is much thinner than the ocean depth and if we neglect magnetospheric plasma effects, the magnitude and phase of the induced field measured by the Galileo spacecraft only depends on the size of the conducting material (ocean's depth $h$), the size of the moon $R_E$, the phase speed $c=\omega R_E$ of the Jovian field, and the electrical properties of the ocean. In the literature, the ratio $\omega \mu_0\sigma R_E^2$ between the skin depth and the moon's radius is generally used to quantify the induction within the ocean. In the present case, we rather introduce the phase speed $c=\omega R_E$ of the magnetic field and use the magnetic Reynolds number $Rm=ch \mu_0 \sigma$, which is more natural  for describing magnetohydrodynamic effects in a channel of thickness $h$. Using this definition, $Rm$ can also be regarded as the ratio between some typical length scale $\sqrt{R_Eh}$ and the skin depth over which electrical currents penetrate the ocean.

Fig.2 has been obtained by performing approximately $100$ DNS with various values of $Ek$, $\Lambda$ and $Pm$, corresponding to different values of $Rm$.  For each of these simulations, we solved Laplace equation outside our spherical domain to compute  the magnetic field beyond the ocean, simulated Galileo spacecraft trajectories and compared to space mission data.During its mission, Galileo made multiple flybys of Europa on several orbits around Jupiter (labeled E4, E11, E12, E14, E15, E17 and E19) if we restrict to encounters for which magnetometer data were acquired. Magnetic measurements from encounters E11, E12, E15, E17, and E19, exhibit short-scale fluctuations probably generated by disturbances due to the plasma currents from the Alfven wings. Following previous authors$^{33}$, we therefore restricted our analysis to the remaining encounters E14 and E4, which are closer to the equatorial plane, outside of the Jovian current sheet. As the two flybys provide the same conclusion (best fit obtained for $\sigma\sim1S/m$), we only show here the E14 flyby. Similarly to previous authors, we use the so-called Ephio coordinate system. Fig.2 was obtained with simulations performed at a fixed value of the aspect ratio (corresponding to an ocean depth of $h=142$ km). Note that  local magnetospheric plasma disturbances, by reducing the time offset between imposed and induced fields, are expected to improve the agreement with observations$^{33}$.

{\bf 3. Simple reduced model}

The behavior observed in our simulations can be understood by studying the equatorial region of Europa, where the magnitude of the oceanic jet is significant. In this region, Jupiter's magnetic field is well approximated by an homogeneous horizontal magnetic field  rotating in the $\phi$-direction at constant speed. For simplicity, let us assume that both induced magnetic field and electrical currents can be regarded as traveling waves propagating in the azimuthal direction, such that the total magnetic field writes:

\begin{equation} 
{\bf B}=(B_r(r){\bf e_r}+B_\phi(r){\bf e_\phi}) e^{i(\phi-\omega t)},
\end{equation}

which includes both Jovian and induced magnetic fields.  As we focus on the equatorial plane, the $\theta$-dependance of the fields is ignored, and induced currents are assumed to be only along the $\theta$-direction. Similarly, we seek for a  simplified velocity field ${\bf U}=U_\phi(r){\bf e_\phi}$. The ocean is confined in the spherical gap between $r=R_i$ and $r=R_i+h=R_E$, where $h$ is the thickness of the ocean. In this reduced model, we also ignore the non-time varying axial component of Jupiter's field, as it would require to seek for more complicated velocity and magnetic fields.

When using the vector potential such that  ${\bf B}={\bf \nabla \times}( A(r)e^{i(\phi-\omega t)}{\bf e_\theta})$ , the induction equation becomes:
  
\begin{equation}
 -i\omega A+\frac{iA}{r\sin\theta}U(r)=\frac{1}{\mu_0\sigma}\left[  \frac{1}{r^2}\frac{\partial (r^2\partial_rA)}{\partial r} - \frac{2A}{r^2\sin^2\theta}  \right] 
\end{equation}

At the boundaries $r=R_E$, the magnetic field $B_\phi$ is required to match the applied Jovian magnetic field, $B_\phi(R_i,R_E)=B_0$, where $B_0$ represents the magnetic field of Jupiter at Europa's location. This boundary condition is similar to what is generally done in such electromagnetically-driven flows \cite{Gailitis76}$^{,34}$. At the lower boundary $r=R_i$, we use $B_\phi(R_i,R_E)=0$, since the time-varying magnetic field wave can not  diffuse into the interior due to the large $Rm$. 
  
In the limit of an ocean thickness small compared to the moon's radius and by focusing on the jet very close to the equatorial plane ($\theta\sim\pi /2$), the induction equation then becomes:

\begin{equation}
\partial_{rr} A=\frac{2}{R^2} \left[1-i\frac{\mu_0\sigma R}{2}\left(c-U(r)\right)\right]A\\
\label{induction_model}
\end{equation}

where  $R=(R_i+R_E)/2$ is the mean radius of the shell and $c=\omega R$ is the phase speed of the traveling magnetic field of Jupiter. This equation is associated with the boundary condition $\partial_rA|_{R_E}=B_0$. 
  Similarly, the electrical currents induced in the channel are given by ${\bf j}=\frac{1}{\mu_0}{\bf \nabla \times B}=-\frac{1}{\mu_0}{\bf \Delta A} $. Using equation (\ref{induction_model}), the electrical currents read :
  
  \begin{equation}
  {\bf j}=i \frac{\sigma A}{R}(c-U(r)){\bf e_\theta} = \sigma(U(r)-c)B_r{\bf e_\theta} 
  \label{currents}
  \end{equation}
  
where $B_r$ is the radial component of the magnetic field. A closed system of equations is obtained by writing the $\phi$-component of the time-averaged Lorentz force:

\begin{equation}
F=-\frac{1}{2}\Re\{j^* B_r\}=\frac{\sigma}{2R^2}(c-U(r))|A(r)|^2
\label{lorentz}
\end{equation}

By using this expression of the Lorentz force in the time-averaged Navier-Stokes equation, one finally obtains the additional equation:
\begin{equation}
\partial_{rr} U= -\frac{\sigma}{2\rho\nu R^2}(c-U(r))|A|^2 ,\label{NS_model}
\end{equation}

{\bf 4. Derivation of the scaling law}

Equations (\ref{induction_model}) and (\ref{NS_model}) can be easily solved numerically and compared to our DNS. However, most of the results reported here can be understood without such a numerical integration. Under the so called 'block velocity' assumption~\cite{Gailitis76}$^{,34}$, we suppose that the velocity $U$ is constant across the gap, except in some boundary layers that we will ignore for now. In this case, equation (\ref{induction_model}) can be integrated:

\begin{equation}
\left[ \partial_rA\right]_{R_i}^{R_o} = \frac{2}{R^2} \left(1-i\frac{\mu_0\sigma R}{2}\left(c-U\right) \right)\int_{R_i}^{R_E}A dr\\
\end{equation}

By using boundary conditions for the potential vector, we obtain an expression for the spatially averaged magnetic field  ${\overline B_r}=-\int iA/(hR) dr$ : 

\begin{equation}
{\overline B_r}=-\frac{1}{h}\int_{R_i}^{R_E}(\frac{iA}{R})dr =  \frac{-iB_0}{2\chi \left(1-i\frac{\mu_0\sigma R}{2}\left(c-U\right) \right)}
\end{equation}
 
 By combining this averaged expression of $B_r$ with  (\ref{currents}) and (\ref{lorentz}), we obtain a simple expression for the Lorentz force acting on the ocean:
 
 \begin{equation}
{\overline {\bf F}}= \frac{\sigma B_0^2c(1-Q)}{8\chi^2+2Rm^2(1-Q)^2}{\bf e_\varphi}
\label{expressionF_full}
\end{equation}

where $\chi=h/R$ is the aspect ratio and $Q=U/c$ is the velocity of the ocean in the midplane normalized by the traveling speed of the Jovian field. We have also introduced the magnetic Reynolds number $Rm=c\mu_0\sigma h$. A simple balance between the Lorentz force and the viscous force therefore suggest :

\begin{equation}
U \propto \frac{\sigma B_0^2c(1-Q)}{\rho\nu\left( 8\chi^2+ 2Rm^2(1-Q)^2\right)}
\end{equation}

Finally, note that in Jovian moons, it is obvious than the ocean always flows with a typical velocity much smaller than the propagation speed of the Jovian field $c=230$m.s$^{-1}$, independently of the source of motions. It is thus reasonable to take the limit $Q=U/c \ll 1$ in the above expression. In addition, Galileo measurements suggest  electrical conductivity around $\sigma\sim 1S/m$, such that the magnetic Reynolds number is much larger than unity. The previous expression therefore simplifies to:

\begin{equation}
U \propto \frac{B_0^2}{\mu_0^2\rho\nu \sigma c}
\label{scaling}
\end{equation}

Alternatively, this expression can also be written in dimensionless form $Q\propto N$, where $N=B_0^2/(\mu_0^2\rho\nu\sigma c^2)$ is a new dimensionless control parameter for the ocean's evolution. As shown in Fig.3a of the main manuscript, all our DNS collapse on this predictive scaling law. Note that the coefficient of proportionality can not be deduced from the reduced model derived here, as it involves details on velocity field fluctuations, geometry of the considered channel, etc.  However, the rescaling of our DNS on a single curve provides an estimate of this coefficient, our data suggesting $Q\sim 0.2N$ (solid black line).

Even if a fairly good agreement between the model and the DNS is obtained here, note that the present model ignores several effects, such as the global rotation of the planet or other driving forces. In addition, as discussed in the main part of the article, the huge turbulent fluctuations expected at large Reynolds number may strongly change the dissipation, such that the $\nu$ must be understood as an eddy viscosity. Because the Ekman number is far too large and the velocity fluctuations remain moderate in our simulations, it is relatively difficult to estimate the value of eddy viscosity in Europa's ocean. However, rotating tank experiments~\cite{Campagne2016} have recently shown that global rotation, by producing a two-dimensionalization of the flow,  strongly reduces friction. Due to Proudman-Taylor constrain, only the $3D$ part of the flow participate to the eddy viscosity $\nu_t\sim C_D U_{3D}H$, leading to values significantly larger than the non-rotating case. Although it is speculative, we argue that such a turbulent viscosity might lead to flows of a few cm/s or so.

\noindent {\bf 5. Comparison with tidally-driven and thermally-driven flows}

{\noindent \bf Tidal effects}: While orbit eccentricity produces flexing of the solid parts of the moon and provides most of the heat that sustains the ocean~\cite{Ross1986}, the flow response to tidal forcing can involve either eccentricity or obliquity. In the first case, the tidal dissipation strongly depends on the ocean thickness $h$, and is associated to time-dependent cellular flows with velocities inversely proportional to $h$, leading to small velocities of a few mm.$s^{-1}$ for Europa. When obliquity of the orbit is taken into account, flow velocities of about $9$ cm$.s^{-1}$ may be generated, although such large flows require resonant excitation of Rossby waves~\cite{Tyler2008}.

{\noindent \bf Thermal buoyancy}: Radiogenic flux of roughly $4$ mW.m$^{-2}$ from the silicate interior also contributes to the heating of the subsurface ocean, and generates hydrothermal plumes ascending from the seafloor. These plumes merely depends on the ration between fluid momentum transport and Coriolis forces, measured by the so-called natural Rossby number $Ro=(Bf^{-3})^{1/4}/h$, in which $B$ is the buoyancy flux emitted by the seafloor source and $f=2\Omega\sin(\theta)$ the Coriolis parameter. While some authors suggested a seafloor heat flux around $Q\sim0.1$ GW$^{35}$, others rather used $Q\sim 1-10$GW\cite{Thomson2001}, leading to Rossby numbers in the range $Ro\sim 0.01 - 0.1$. The  corresponding velocities associated with hydrothermal plumes are predicted~\cite{Goodman2004,Goodman2012} to lie between $0.9$cm$.s^{-1}$ and $5$cm$.s^{-1}$. In a more recent paper\cite{Soderlund2013}, it was suggested that Rossby number may reach even larger values, $Ro>1$. In this extreme case, one may expect quasi-3D turbulent convection in the ocean producing retrograde zonal flows of about $250$cm$.s^{-1}$, comparable to the velocities produced by our MHD mechanism. 

In conclusion, the magnetically-driven jet, with velocities comprised between $1$ and $100$ cm$.^{-1}$,  is a non-negligible source of motions in Europa's ocean. Moreover, if the convective Rossby number is smaller than one, this may be  the only contribution to the time-averaged axisymmetric zonal flow, which may significantly affects the non-synchronous rotation of the moon.

\noindent {\bf 6. Numerical simulations including thermal buoyancy}

The above discussion suggests that the mechanical flow response (of a few mm$.s^{-1}$) to tidal forces may be neglected, whereas hydrothermal plumes due to radiogenic heating from the mantle remain crucial. Therefore, a full numerical model of Jovian moons should at least combine magnetohydrodynamical effects and thermal buoyancy due to radiogenic heating. 

 In some of our DNS, previous MHD equations have therefore been coupled to thermal Boussinesq convection with fixed isothermal boundary conditions, introducing the Rayleigh number $Ra=\alpha g \Delta TR^3/(\nu\kappa)$ and the thermal Prandtl number $Pr=\nu/\kappa$, where $\Delta T$ is the unstable superadiabatic gradient imposed through the ocean, $\kappa$ is the thermal diffusivity, $g$ is the gravity,  and $\alpha$ is the thermal expansion coefficient. We can also define a convective Rossby number $Ro_c=\sqrt{RaE^2/Pr}$, similar to the natural Rossby number but based on the temperature contrast. Because a complete thermal model of Europa is beyond the scope of the present paper, we only discuss the effect of a thermal gradient imposed between inner and outer spheres and ignore internal heat source as produced by tidal heating.

Typical values for Europa have been estimated by various authors\cite{Soderlund2013,Goodman2004}, suggesting $Pr\sim 12$ and $Ra\sim 10^{20}-10^{23}$, corresponding to convective Rossby numbers between $0.01$ and $1$. As usual, numerical limitations do not allow us to reproduce the low viscosity of Europa's ocean. We present results obtained for larger values of Ekman and smaller Rayleigh numbers, $E=10^{-5}$ and Rayleigh numbers ranging from $Ra\sim 10^{7}$ to $Ra\sim 10^{10}$, such that we reproduce the range of Rossby numbers,  $Ro \sim 0.01-1$ relevant to Europa.

Supplementary Figure $1$ shows snapshots of the azimuthal velocity field at mid-height from the seafloor obtained for $E=10^{-5}$, $\Lambda=10^{-1}$ and $Pm=10^{-3}$ and $Pr=12$,  for $5$ different values of the Rayleigh number. For numerical convenience, note that contrary to Fig. 4, this set of parameters corresponds to the upper bound of the magnitude of the jet (viscous scaling), although the same conclusion can be drawn for both regimes. It shows an increase of small scale convection patterns as the Rayleigh number is increased (from quasi-2D columnar plumes to 3D convection). At small Rossby number ($Ro<1$), thermal convection only weakly modifies the large scale average structure of the MHD jet. Interestingly, for $Ro>0.4$, thermal convection also generates a retrograde zonal flow near the equatorial region of the outer sphere, which even reinforces the magnetically-driven jet.

 As long as the convective Rossby number is smaller than unity, panel (f) shows that the large scale flow is always dominated by the magnetically-induced retrograde jet. Because the Rossby number for Europa is believed to lie in this limit $Ro<1$, most of the results reported in the paper are almost unaffected by the presence of thermal buoyancy.  Note however  that our simulations predict that the small scale part of the flow may be dominated by rotating thermal plumes with typical velocities of a few cm$.s^{-1}$. Finally, if the convective Rossby number is as large as suggested by Soderlund et al\cite{Soderlund2013} ($Ro\sim 1$), a retrograde jet of several hundreds of $cm/s$ is generated, nearly three times larger than the magnetically-driven jet.  Similarly, supplementary figure 2 shows that Joule heating is only weakly modified by the presence of thermal convection.

\begin{addendum}
\item[Data availability] The data that support the plots within this paper and other findings of this study are available from the corresponding author C. Gissinger upon reasonable request.
\end{addendum}

\noindent 31. Christensen U., et al  {\it PEPI}, {\bf 128} (2001)

\noindent 32. Dormy E., Cardin P. and Jault D., {\it Earth Planet. Sci. Lett.}, {\bf 160}, 15, (1998)

\noindent 33. Zimmer, C., Khurana, K.K., Kivelson, M.G. Subsurface oceans on Europa and Callisto: Constraints from Galileo magnetometer observations. {\it Icarus} {\bf 147}, 329-347 (2000)

\noindent 34. Gissinger, C., Rodriguez-Imazio, P., Fauve, S., {\it Instabilities in electromagnetically-driven flows, part I}, {\it Phys. Fluids} {\bf 28}, 034101 (2016)

\noindent 35. Vance, S., Goodman, J.C. Oceanography of an ice-covered Moon. In: Pappalardo, R.T., McKinnon, W.B., Khurana, K.K. (Eds.), Europa. University of Arizona Press (2009)

\end{methods}

\newpage

\end{document}